\documentclass[submission,copyright,creativecommons]{eptcs}
\providecommand{\event}{WS-FMDS 2012} 
\usepackage{breakurl}             
\usepackage{graphicx}

\title{ROSA Analyser: An automatized approach to analyse processes of ROSA \thanks{This work has been partially supported by projects TIN2009-14312-C02-02 \& CGL2010-20787-C02-02}}

\long\def\comment#1{}

\def\dissumo{ \sum \kern -1.30em \odot \kern 0.6em}

\newcommand{\excho}{\,+\,}

\def\flechadcha{> \kern -0.25em \longrightarrow}

\newcommand{\incho}{\,\oplus\,}

\def\M1{{\bf M_1}}

\def\sumo{ \sum \kern -1.05em \circ \kern 0.60em}

\newcommand{\prcho}{\oplus}

\newcommand{\bex}{\begin{example} \begin{rm}}
\newcommand{\eex}{\vspace{-4ex}{\flushright $\Box$\\ \mbox{}\vspace{-4ex}} \end{rm} \end{example}}

\author{Ra\'{u}l Pardo and Fernando L. Pelayo
\institute{Dept. de Sistemas Inform\'{a}ticos\\
E. Superior de Ingenier\'{i}a Inform\'{a}tica de Albacete\\
Universidad de Castilla - La Mancha\\
Campus Universitario. 02071-Albacete, Spain\\}
\email{raulpardo@dsi.uclm.es, FernandoL.Pelayo@uclm.es}}

\def\titlerunning{An automatized approach for analysing processes of ROSA}
\def\authorrunning{Ra\'{u}l Pardo \& Fernando L. Pelayo}
\begin{document}
\maketitle

\begin{abstract}

In this work we present the first version of ROSA Analyser, a tool 
designed to get closer to a fully automatic process of analysing 
the behaviour of a system specified as a process of the Markovian 
Process Algebra ROSA. In this first development stage, ROSA 
Analyser is able to generate the Labelled Transition System, 
according to ROSA Operational Semantics. 

ROSA Analyser performance starts with the Syntactic Analysis so 
generating a layered structure, suitable to then, apply the 
Operational Semantics Transition rules in the easier way. ROSA 
Analyser is able to recognize some states identities deeper than 
the Syntactic ones. This is the very first step in the way to 
reduce the size of the LTS and then to avoid the state 
explosion problem, so making this task more tractable. 

For the sake of better illustrating the usefulness of ROSA 
Analyser, a case study is also provided within this work. 

\end{abstract}

\section{Introduction}

Formal methods are more used as Computer Science becomes a more 
mature science; this happens due to the fact that formal methods 
provide software designers with a way to guarantee high security 
and reliability levels for their computer science systems, and 
what is more, formal methods would allow to find software errors 
in the earliest stage of the software development life cycle so 
making it significatively cheaper. The main problem is that for 
systems with a size or complexity level, the analysis could be 
difficult to be developed, mainly because of two reasons, the 
first comes from the size of the graphical model usually 
generated, the second is because the most times these analyses are 
made by hand; we begin with the latter so presenting a tool to apply
automatically the operational semantics of the Process 
Algebra, ROSA. In the next future we will continue dealing with 
the problem of the size of the LTS generated by ROSA. 

During the past 20 years many efforts have been done in order to 
develop support tools for all kinds of formal methods. For 
instance for timed automata UPPAAL tool \cite{UPPAAL96} gives to 
designers a good environment to design using timed automata, 
as well as, it supports analyzing skills. In the Petri Nets field, 
TINA is a frequently used tool \cite{TINA04}. Regarding Process 
Algebras, several tools have been developed also, the most known 
could be PEPA Workbench \cite{GH94} which was developed based on 
PEPA process algebra \cite{H96}. The main problem of process 
algebra based tools is to deal  with the \emph{state explosion} arisen in the Labelled Transition System (LTS), because of 
this, some of the process algebra research lines are searching the 
way to reduce the size of the LTS to be generated, and so to avoid 
this problem as much as possible \cite{SHB11}. 

In this paper we present a tool based on ROSA process algebra 
\cite{Pelayo04}, so this tool has to deal with non-deterministic, 
probabilistic and temporal behaviours for a given process, which 
means that it has to provide a proper syntax for capturing all 
these features. Moreover, an easy and clear user interface has 
been designed for it which has been not very common in previously 
developed tools. 

In its first development stage ROSA Analyser is able to build the 
LTS from a given ROSA process, by means of applying all of the 
possible operational Semantics rules which must be used for a 
given input process. In addition, for the sake of avoiding the 
\emph{state explosion} problem, we are working on defining an 
heuristic which allows to reach from each state/process the most 
probably path to a given (final) state/process, by changing its 
exponential computational cost to a linear computational cost. 
Therefore, it involves an initial attempt to solve this problem 
and as a consequence of this, get a more practical way to analyse 
concurrent and real time systems by means of process algebra based 
tools. 

The paper is structured as follows, section 2 provides a detailed 
description of ROSA Analyser, pointing out the type of used data 
structures and the analyzing process which makes possible the LTS 
building; in section 3 a case of use of it is presented, 
specifically a cognitive memorizing process has been analysed. 
To finish with, section 4 states both conclusions and future work. 

\section{ROSA Analyser tool}

ROSA Analyser has been developed in JAVA, this choice was taken 
due to the facilities to handle with complex data structures, by 
which is characterized this programming language. In addition, 
ROSA Analyser is implemented over GPL v2 license, in order to 
allow everybody the possibility to work on the defined data 
structures. The source code and the executable version can be 
found in \url{http://kenai.com/projects/semantictreerosa}. 

In order to show the way in which ROSA Analyser works, we will 
describe all of the main parts involved in the LTS generator. 
Below paragraphs detail the data structures over which ROSA Analyser works, then the 
Semantics analyser, containing the main work load, is executed so producing a kind of layered structure
appropriate to be used by the latter process which builds the whole LTS form the 
original syntax expression of the input ROSA process. 

\subsection{Data structures}

The theoretical process expressions with which ROSA Analyser works 
are not optimized for the analysis in which the tool has to be 
involved, due to this fact, the data are needed to be parsed in order to
reach a better structure for their analysis processes. Specifically, in 
this section we show how the tool handle the data structures required
for the Syntax and Semantics analysis. 

\subsubsection{Syntax Tree}

As a result of the syntax analysis a binary tree is built with 
which can be made an easy and efficient Semantics analysis. In 
this Syntax analysis the higher syntactical priority the elements 
have, the closer to the top they are located, as previously said 
this structure is optimized to check the conditions of the upper 
side of the rules and to do the necessary modifications to the 
process which is being analysed. 

\begin{figure}[!hbtp]
    \begin{center}
        \includegraphics[scale=0.5]{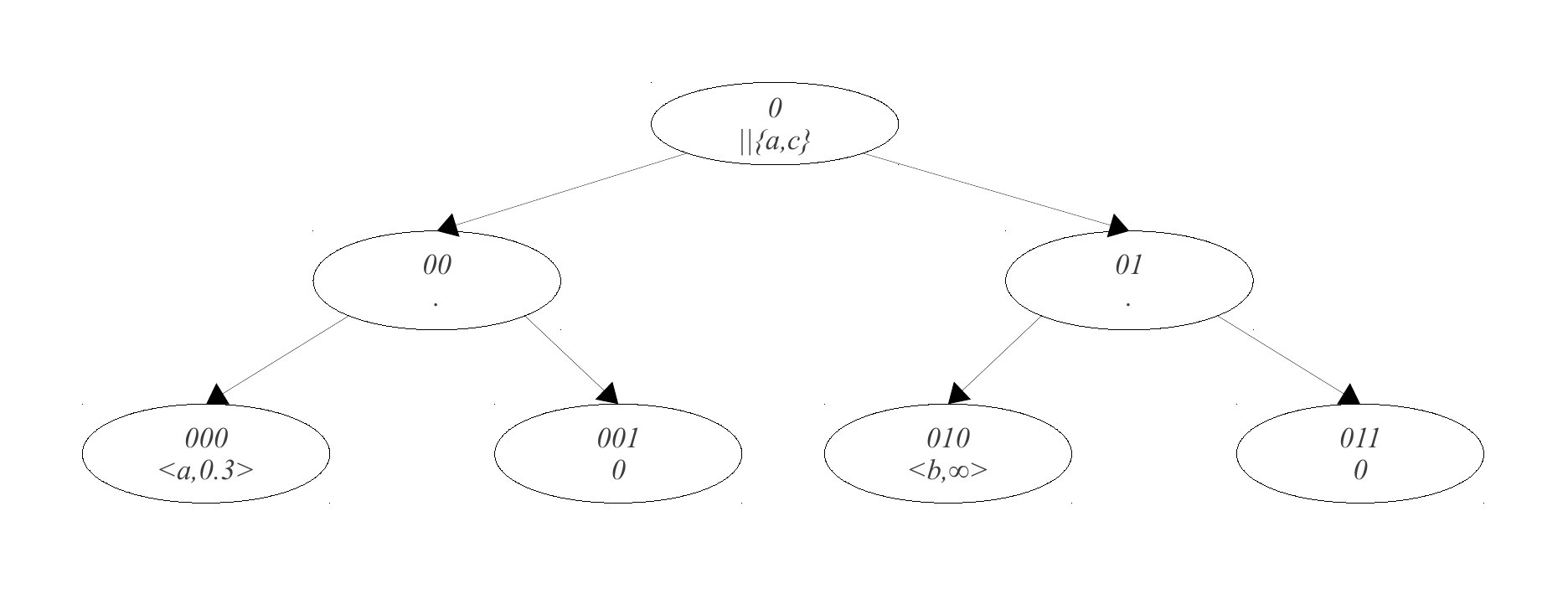}  
    \end{center}
    \caption{Syntax Tree Example}
    \label{SyntaxTreeEx}
\end{figure}

In order to show in a clear way this data structure 
an example of the syntax tree of the ROSA process $<a,0.3>.0 
\vert\vert_{\{a,c\}} <b,\infty>.0$ can be seen in figure \ref{SyntaxTreeEx}. 

In this example, we can see the root node corresponded with
the highest priority element of the process, also this structure
allows the semantics analysis to get directly the element which
chooses the rule to be applied, and by means of defined methods 
also make possible change the Syntax tree so as to apply the 
selected rule. Detailed description about the building of the Syntrax tree will be provided.

\subsubsection{Semantics Tree}


Once we have defined a proper structure for the process' syntax we 
are ready to show the Semantics tree data structure which 
represents the LTS for a ROSA process. The LTS is built with nodes 
which are themselves ROSA processes, therefore in each Semantics 
tree node will appear a Syntax tree, this can show the high 
complexity by which is characterized this data structure. Also, 
unlike the Syntax tree, the transitions between nodes must get 
more information than those of the Syntax tree which further 
increase the amount of data to be handle. 

\begin{figure}[!hbtp]
    \begin{center}
        \includegraphics[scale=0.4]{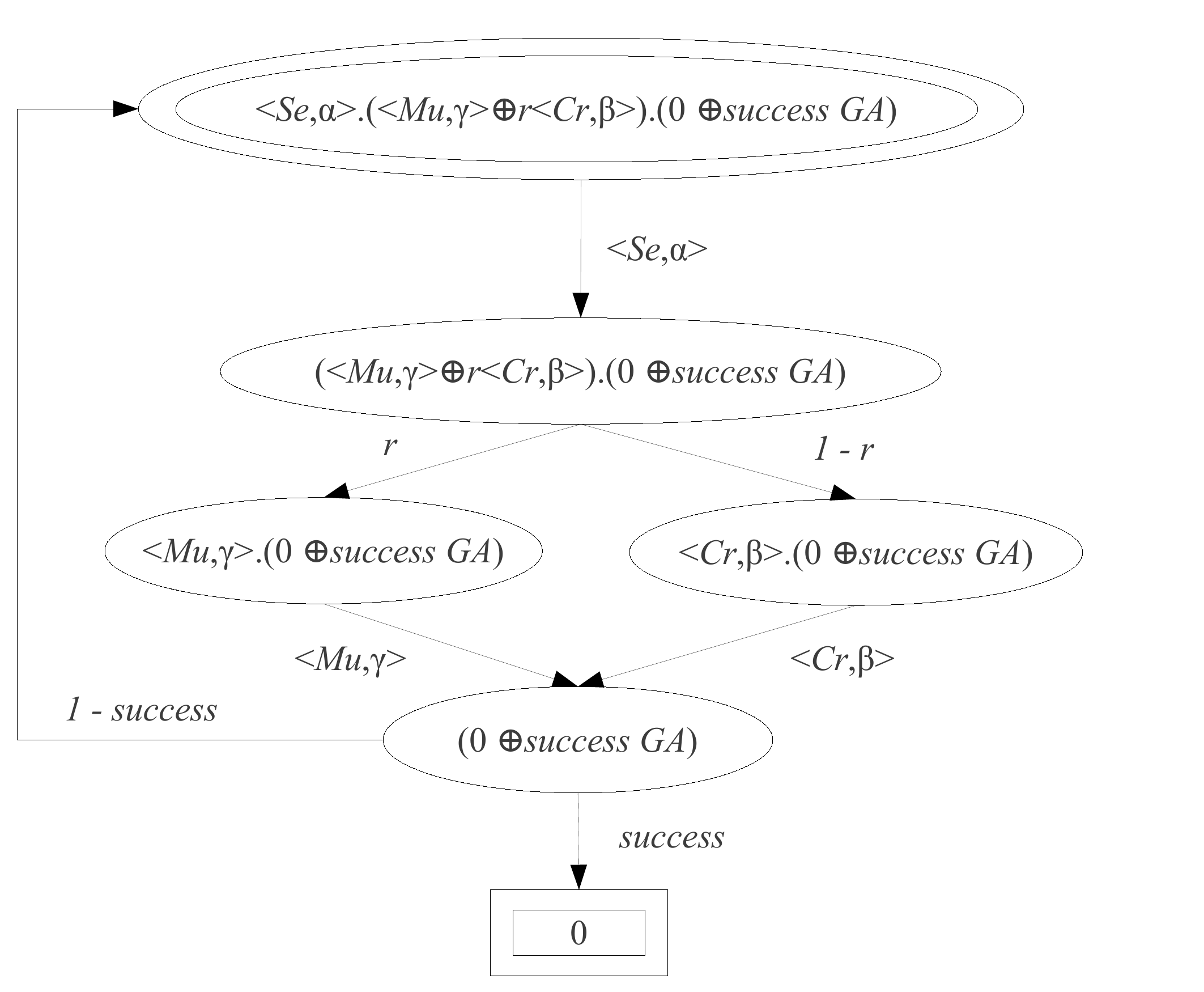}   
    \end{center}
    \caption{Semantics Tree Example}
    \label{SemanticsTreeEx}
\end{figure}

In this case the defined tree structure is chosen with the aim to 
support the natural structure of the Semantics tree, which is a 
n-ario tree, unlike the previous data structure which was chosen 
looking for making easier the Semantics analysis. A Semantics tree 
example can be seen in figure \ref{SemanticsTreeEx}. 

It is important to point out that this data structure also has 
several methods in which some functionalities which help to build the semantics tree
have been implemented , such as joining new nodes to the 
Semantics tree, where in addition, a search for Syntax equivalent 
nodes is made everytime that a tentative ``new" node is 
generated. 

\subsection{Syntax Analysis}

\subsubsection{Tool input Syntax}
\paragraph{}
The Syntax analysis consists of the part of the tool which takes 
as input the plain expression of a ROSA process and returns the 
Syntax tree associated with that process. It was necessary to 
define some constrains for the input syntax because the ASCII 
characters do not provide the wide variety of elements by which is 
characterized ROSA Syntax. In the following table \ref{EquivSyn} 
are shown the equivalences between ROSA Syntax and the tool 
input Syntax: 

\begin{table}[!hbtp]
\begin{center}
\begin{tabular}{|c | c|}
\hline
\textbf{ROSA Syntax} & \textbf{Tool Syntax}\\
\hline
$ a $ & $ a $\\
$ <a,\alpha> $ & $ <a,\alpha> $\\
$ a.P $ & $ a.P $\\
$ <a,\alpha>.P $ & $ <a,\alpha>.P $\\
$ P $ & $ P $\\
$ P;Q $ & $ P;Q $\\
$ P \incho Q $ & $ P - Q $\\
$ P \excho Q $ & $ P + Q $\\
$ P \prcho_{r} Q $ & $ P *\{r\} Q $\\
$ P \vert \vert_A Q $ & $ P \vert \vert A $ $ Q $\\
\hline
\end{tabular}
\end{center}
\caption{Equivalence between ROSA and tool Syntax}
\label{EquivSyn}
\end{table}

Moreover, it is important to point out that the parameter $\alpha$ 
capturing the temporal behaviour of actions must be a real number, 
and the parameter $r$, a probability, must be a real number within 
the interval $[0,1]$. The set $A$ represents the synchronization 
set for a parallel operator, so that this set is 
$A=\{a,b,c,\ldots\}$. 

\subsubsection{Syntax tree building}

Once the ROSA Analyser input method has been defined we are ready 
to describe the Syntax analysis, the first point to take into 
account is the priority of the operators, by default the priority 
is defined as follows: 

\begin{enumerate}
    \item Sequential processes operator.
    \item Parallel operator.
    \item External, internal and probabilistic choices.
    \item Prefix.
    \item Actions and process variables.
\end{enumerate}

These priorities can be changed according to our needs by means of 
using parenthesis, according to their common behaviour, becoming a
more flexible Syntax for the input method. 

Taking into account the priorities above defined, the Syntax 
analysis is mainly a recursive method in which the expression 
element with the highest priority must be found, then a node with 
the element is created or added to the Syntax tree, once reached 
this point if the element is an operator, the method splits the 
expression into two parts taking as split point the position in 
which was located the operator found, then the same method is 
called using both expressions, and the new elements found in these 
recursive calls are added to the Syntax tree which is being built, 
on the other hand, if not, i.e. if the element with the highest 
priority is an action or a process variable this element is added 
as a node to the Syntax tree but the recursive calls finish. 

\subsection{Semantics Analysis}


Through this Semantics analysis is built the LTS for a ROSA 
process, the complexity of this analysis is very big since it has 
to determine which for all of the ROSA Semantics rules can must be 
used in order to show all of the possible behaviours that a given 
process can perform. 

Firstly, a LTS constituted with a unique Semantics node which 
contains the Syntax tree build for the input process is created, 
this node will be the root node in the Semantics tree. In the next 
step it has to be determined which rule for all of the ROSA 
operational Semantics rules can be applied. As it can be seen in 
ROSA operational Semantics there are three rule sets 
(non-deterministic, probabilistic and action transition rules). 
Then the first checking in which ROSA Analysis is involved is to 
determine whether the process is non-deterministic by means of 
\emph{deterministic stability} function, if the process is 
non-deterministic it will be checked which of the 
non-deterministic transition rules must be used, that will be 
checked by means of the root node in the Syntax tree. 

\begin{figure}[!hbtp]
    \begin{center}
        \includegraphics[width=130mm]{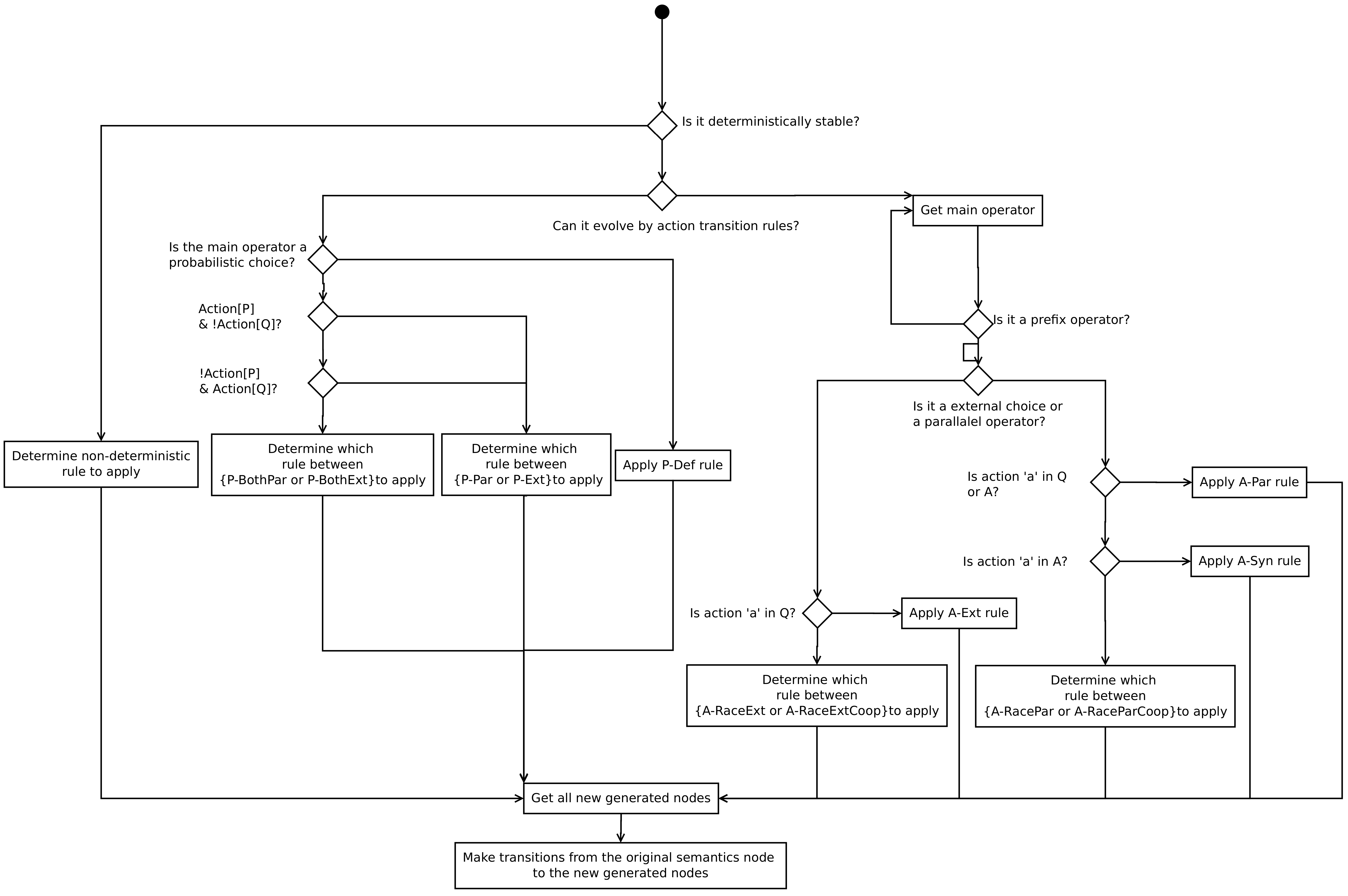}
    \end{center}
    \caption{Activity diagram for Semantics nodes analysis process}
    \label{NodeAnalysis}
\end{figure}

On the other hand if the process is deterministically stable, the 
tool must determine trough the \emph{action} function if the 
process can be evolved by probabilistic or action transition 
rules. Once the transition rules set has been chosen, ROSA 
Analyser determines, by means of the upper condition, which rule 
must be applied. In order to show a clearly way of this process, 
figure \ref{NodeAnalysis} shows its activity diagram. 

Once all these nodes have been created, they must be joined to the 
LTS. As it is well-known the \emph{state explosion} problem 
arises when the LTS is being built, but this tool, in order to 
reach a lighter solution for this problem, is able to detect 
Syntax identities between nodes in order to avoid a node to appear 
more than once. When the new nodes have been added to the LTS, 
again all leaf nodes will be analysed, in order to search more 
nodes that could be generated. This loop is repeated until two 
consecutive iterations do not provide any new node, this will mean 
that the LTS is completed. 

\section{Analysing a memorizing process}

In order to show the usefulness of this tool, in this section we 
describe the way in which a memorizing process is analysed. This 
process was specified using ROSA in \cite{PPCVD07}. 

Now we are going to describe the input process and the 
equivalences with the original specification. However, to get a 
clearer description we roughly describe each process involved and 
its corresponding ROSA specification. 

\subsection*{Encoding process}

This is the process with which the memorizing process starts, 
firstly elements identification is computed, once this initial 
identification is complete then at the same time it tries to 
identify objects and relations with previous data. The ROSA 
specification for this process can be seen in 
(\ref{EncodingROSA}): 

$$
Encoding \equiv <Perception,\beta>.<Start\_Identification,\infty>.(<Find\_Attributes,\alpha_1>.<stm,\infty>\vert\vert_{\{stm,abm\}}
$$
\begin{equation}
\vert\vert_{\{stm,abm\}}<Identify\_Object,\alpha_2>.<stm,\infty>\vert\vert_{\{stm,abm\}} 
<Find\_Relations,\alpha_3>.<stm,\infty>) \label{EncodingROSA} 
\end{equation}
\\
The specification used as input for the tool is:
\begin{equation}
E \equiv 
<a,0.1>.b.(<c,0.2>.f\vert\vert\{f,i\}<d,0.3>.f\vert\vert\{f,i\} 
<e,0.4>.f) \label{EncodingTOOL} 
\end{equation}

\subsection*{Consolidation process}

The memorizing process specified has three stages for memory 
storage: \emph{Sensory store}, \emph{Short term memory} and 
\emph{Long term memory}. The first two kinds of memory aren't 
definitive, i.e. not consolidated yet, and the information 
perceived is stored several seconds there, but when the 
information is longer stored, we are consolidating this 
information. Because of this, once the encoding process has 
finished with a certain amount of probability (according to 
parameter $ r $ in full memorizing process specification) the 
perceived information can be stored in long term memory or what is 
the same, it is definitively consolidated. The ROSA specification 
for this process is: 

\begin{equation}
    Consolidation \equiv <Soon\_Rehearsal\_or\_Finding\_Cues, \alpha_4>.<ltm,\infty>.<abm,\gamma>
\end{equation}
\\
In the same way that the previous process was used, the 
specification for ROSA Analyser is: 

\begin{equation}
    C \equiv <g,0.5>.h.<i,0.6>
\end{equation}

\subsection*{Retrieval process}

Retrieval process represents the way in which the memorizing 
process access to information which has been previously stored, 
ROSA specification for this process is: 

\begin{equation}
    Retrieval \equiv <Request,\infty>.<abm,\sigma>
\end{equation}
\\
Therefore, as ROSA Analyser input we will use:
\begin{equation}
    R \equiv j.<i,0.7>
\end{equation}

\subsection*{Losing process}

With a certain probability it is possible to lose the perceived 
information, this event can happen in the case the 
information does not reach the long term memory. Obviously, this 
process only represents a single action execution, so in the 
specification there is only one action involved: 

\begin{equation}
    Losing \equiv <Fading,\alpha_5>
\end{equation}

For tool Syntax we take:
\begin{equation}
    L \equiv <k,0.8>
\end{equation}

\subsection*{LTS for the memorizing process}
Finally, joining all of the defined process the expression 
associated for the memorizing process is: 

\begin{equation}
    Mem \equiv Encoding;(Consolidation \prcho_{0.25} Losing)\vert\vert_{\{abm\}} Retrieval
\end{equation}
 \\
 And for ROSA Analyser this process must be written as follows:
\begin{equation}
    M \equiv E;(C*\{0.25\}L)\vert\vert\{i\}R
\end{equation}

Once the Semantics analysis has been performed, ROSA Analyser 
provides two outputs, one in which the nodes and transitions can be 
textually seen, and other graphical output that by means of 
graphviz the LTS, can be exported to pdf, jpg, svg and eps. In 
url \url{http://raulpardo.files.wordpress.com/2012/05/m.jpg} we can see the latter. 

As a matter of proof of usefulness of ROSA Analyser, this LTS has 
two distinguished states, one red coloured representing a deadlock 
state, and a green coloured state representing a success. 

\section{Conclusions and future work}

To sump up we can say that a tool to automatize the LTS building 
process of a ROSA process has been developed. This tool parses the 
process Syntax expression into a layered data structure quite 
suitable to better develop the Semantics analysis, such 
Operational Semantics provides the rules among which a proper 
subset of them, for each process should be applied. As a matter of 
fact, we have shown a real example by which the practical use of 
ROSA Analyser can be seen. In addition, it is important to mention 
that this tool provides two kinds of outputs, a textual output in 
which all nodes and transition can be seen; and a graphical view 
in which the Semantics tree is drawn in a portable format. 



Although we reach a more practical use of ROSA process algebra, 
the \emph{state explosion} problem has not been resolved yet. 
At this moment, ROSA Analyser is able to ``see" more identities of 
processes than the Syntactical one so reaching a little reduction 
in the states number. However, as we previously stated some first 
steps in this line have been done, specifically, a metric over the 
set of ROSA processes and a sort of normal form has been defined 
in \cite{PCC01}, through this distance and normal forms can be 
defined several heuristics which provide the tentative more 
promising branch from any state to a given final one, so 
decreasing the number of states and making this task more 
practically tractable.


\bibliographystyle{eptcs}

\begin{thebibliography}{1}
\providecommand{\bibitemdeclare}[2]{}
\providecommand{\surnamestart}{}
\providecommand{\surnameend}{}
\providecommand{\urlprefix}{Available at }
\providecommand{\url}[1]{\texttt{#1}}
\providecommand{\href}[2]{\texttt{#2}}
\providecommand{\urlalt}[2]{\href{#1}{#2}}
\providecommand{\doi}[1]{doi:\urlalt{http://dx.doi.org/#1}{#1}}
\providecommand{\bibinfo}[2]{#2}

\bibitemdeclare{incollection}{UPPAAL96}
\bibitem{UPPAAL96}
\bibinfo{author}{Johan \surnamestart Bengtsson\surnameend},
  \bibinfo{author}{Kim \surnamestart Larsen\surnameend},
  \bibinfo{author}{Fredrik \surnamestart Larsson\surnameend},
  \bibinfo{author}{Paul \surnamestart Pettersson\surnameend} \&
  \bibinfo{author}{Wang \surnamestart Yi\surnameend} (\bibinfo{year}{1996}):
  \emph{\bibinfo{title}{UPPAAL - a tool suite for automatic verification of
  real-time systems}}.
\newblock In \bibinfo{editor}{Rajeev \surnamestart Alur\surnameend},
  \bibinfo{editor}{Thomas \surnamestart Henzinger\surnameend} \&
  \bibinfo{editor}{Eduardo \surnamestart Sontag\surnameend}, editors: {\sl
  \bibinfo{booktitle}{Hybrid Systems III}}, {\sl \bibinfo{series}{Lecture Notes
  in Computer Science}} \bibinfo{volume}{1066}, \bibinfo{publisher}{Springer
  Berlin / Heidelberg}, pp. \bibinfo{pages}{232--243}.
\newblock \urlprefix\url{http://dx.doi.org/10.1007/BFb0020949}.
\newblock \bibinfo{note}{10.1007/BFb0020949}.

\bibitemdeclare{article}{TINA04}
\bibitem{TINA04}
\bibinfo{author}{B.~\surnamestart Berthomieu~*\surnameend},
  \bibinfo{author}{P.-O. \surnamestart Ribet\surnameend} \&
  \bibinfo{author}{F.~\surnamestart Vernadat\surnameend}
  (\bibinfo{year}{2004}): \emph{\bibinfo{title}{The tool TINA - Construction of
  abstract state spaces for petri nets and time petri nets}}.
\newblock {\sl \bibinfo{journal}{International Journal of Production Research}}
  \bibinfo{volume}{42}(\bibinfo{number}{14}), pp. \bibinfo{pages}{2741--2756},
  \doi{10.1080/00207540412331312688}.
  
\bibitemdeclare{inproceedings}{GH94}
\bibitem{GH94}
\bibinfo{author}{S.~\surnamestart Gilmore\surnameend} \&
  \bibinfo{author}{J.~\surnamestart Hillston\surnameend}
  (\bibinfo{year}{1994}): \emph{\bibinfo{title}{{The {PEPA} Workbench: A Tool
  to Support a Process Algebra-based Approach to Performance Modelling}}}.
\newblock In: {\sl \bibinfo{booktitle}{Proceedings of the Seventh International
  Conference on Modelling Techniques and Tools for Computer Performance
  Evaluation}}, {\sl \bibinfo{series}{Lecture Notes in Computer Science}}
  \bibinfo{volume}{794}, \bibinfo{publisher}{Springer-Verlag},
  \bibinfo{address}{Vienna}, pp. \bibinfo{pages}{353--368},
        \doi{10.1007/3-540-58021-2\_20}.
\bibitemdeclare{book}{H96}
\bibitem{H96}
\bibinfo{author}{Jane \surnamestart Hillston\surnameend}
  (\bibinfo{year}{1996}): \emph{\bibinfo{title}{A Compositional Approach to
  Performance Modelling}}.
\newblock \bibinfo{publisher}{Cambridge University Press},
        \doi{10.1017/CBO9780511569951}.
\bibitemdeclare{phdthesis}{Pelayo04}
\bibitem{Pelayo04}
\bibinfo{author}{F.~L. \surnamestart Pelayo\surnameend} (\bibinfo{year}{2004}):
  \emph{\bibinfo{title}{Application of formal methods to performance
  evaluation}}.
\newblock Ph.D. thesis, \bibinfo{school}{Universidad de Castilla - La Mancha}.

\bibitemdeclare{inproceedings}{PCC01}
\bibitem{PCC01}
\bibinfo{author}{Fernando~L. \surnamestart Pelayo\surnameend},
  \bibinfo{author}{Fernando \surnamestart Cuartero\surnameend} \&
  \bibinfo{author}{Diego \surnamestart Cazorla\surnameend}
  (\bibinfo{year}{2011}): \emph{\bibinfo{title}{Looking for a cheaper ROSA}}.
\newblock In: {\sl \bibinfo{booktitle}{Proceedings of the 11th international
  conference on Artificial neural networks conference on Advances in
  computational intelligence - Volume Part II}}, \bibinfo{series}{IWANN'11},
  \bibinfo{publisher}{Springer-Verlag}, \bibinfo{address}{Berlin, Heidelberg},
  pp. \bibinfo{pages}{380--387}.
\newblock \urlprefix\url{http://dl.acm.org/citation.cfm?id=2023332.2023387}.

\bibitemdeclare{inproceedings}{PPCVD07}
\bibitem{PPCVD07}
\bibinfo{author}{Maria~L. \surnamestart Pelayo\surnameend},
  \bibinfo{author}{Fernando~L. \surnamestart Pelayo\surnameend},
  \bibinfo{author}{Fernando \surnamestart Cuartero\surnameend},
  \bibinfo{author}{Valentin \surnamestart Valero\surnameend},
  \bibinfo{author}{Gregorio \surnamestart Diaz\surnameend} \&
  \bibinfo{author}{Elena \surnamestart Nieto\surnameend}
  (\bibinfo{year}{2007}): \emph{\bibinfo{title}{Does ROSA provide a good view
  of the Memorizing Process?}}
\newblock In: {\sl \bibinfo{booktitle}{Proceedings of the 6th IEEE
  International Conference on Cognitive Informatics}}, \bibinfo{series}{COGINF
  '07}, \bibinfo{publisher}{IEEE Computer Society},
  \bibinfo{address}{Washington, DC, USA}, pp. \bibinfo{pages}{273--283},
  \doi{10.1109/COGINF.2007.4341900}.


\bibitemdeclare{inproceedings}{SHB11}
\bibitem{SHB11}
\bibinfo{author}{A.~\surnamestart Stefanek\surnameend}, \bibinfo{author}{R.A.
  \surnamestart Hayden\surnameend} \& \bibinfo{author}{J.T. \surnamestart
  Bradley\surnameend} (\bibinfo{year}{2011}): \emph{\bibinfo{title}{GPA - A
  Tool for Fluid Scalability Analysis of Massively Parallel Systems}}.
\newblock In: {\sl \bibinfo{booktitle}{Quantitative Evaluation of Systems
  (QEST), 2011 Eighth International Conference on}}, pp.
  \bibinfo{pages}{147--148}, \doi{10.1109/QEST.2011.26}.

\end{thebibliography}

\end{document}